\documentclass[onecolumn,amsmath,amssymb,12pt,nofootinbib]{revtex4}
\usepackage{graphicx}

\newcommand{\nn}{\nonumber}
\newcommand{\be}{\begin{equation}}
\newcommand{\ee}{\end{equation}}
\newcommand{\bea}{\begin{eqnarray}}
\newcommand{\eea}{\end{eqnarray}}

\def\a{\alpha}

\def\d{\delta}

\def\s{\sigma}

\def\e{\epsilon}

\def\f{\phi}

\def\k{\kappa}

\def\th{\theta}

\def\s{\sigma}

\def\t{\tau}

\def\z{\zeta}

\def\pt{\partial}

\def\ie{{\it i.e. }}

\begin{document}

\title{Cosmic Bounces and Cyclic Universes}

\author{
Jean-Luc Lehners
}

\affiliation{\small Max-Planck-Institute for Gravitational Physics (Albert-Einstein-Institute) \\
      D-14476 Potsdam/Golm, Germany }

\begin{abstract}
Cosmological models involving a bounce from a contracting to an expanding universe can address the standard cosmological puzzles and generate ``primordial'' density perturbations without the need for inflation. Some such models, in particular the ekpyrotic and cyclic models that we focus on, fit rather naturally into string theory. We discuss a number of topics related to these models: the reasoning that leads to the ekpyrotic phase, the predictions for upcoming observations, the differences between singular and non-singular models of the bounce as well as the predictive and explanatory power offered by these models.
\end{abstract}

\maketitle

\section{Introduction}

Our universe is special in that it evolved from an {\it a priori} highly unlikely state, in the sense of Boltzmann: the very early universe was exceptionally flat, homogeneous and isotropic (hence it had very low entropy), with approximately scale-invariant density perturbations of magnitude $Q \sim 10^{-5}$ added on top. Now it may be that string theory, once better understood, leads to a unique initial state with these characteristics, but there is currently little evidence pointing in that direction. Therefore we would like to find a {\it dynamical} explanation for these ``initial'' conditions.

The most popular attempt at formulating such a model is inflationary cosmology \cite{Guth:1980zm,Linde:1981mu,Albrecht:1982wi}. In recent years, many attempts have been made at obtaining inflationary dynamics in string theory (see for example the contributions by McAllister and Burgess, Quevedo and Mazumdar to this focus issue). All models to date require rather elaborate constructions (with multiple branes of diverse dimensionalities, orientifold planes and warped geometries induced by $p$-form fluxes all carefully balanced) together with their own set of initial conditions. Thus, currently, these models simply report the question of initial conditions to an earlier epoch, perhaps alleviating it, but not solving it. What these models do achieve is that they provide a mechanism with a microphysical basis for generating the primordial density perturbations, and this is their main merit. All models to date still have to be fine-tuned in order for the characteristics of the generated perturbations, such as their amplitude and spectrum, to agree with observations (see also the contribution by Mulryne and Ward \cite{Mulryne:2011ja}). This unfortunately reduces the explanatory power of inflationary models. A recourse that is sometimes advocated is to invoke eternal inflation \cite{Guth:2007ng} (which creates an infinite number of universes of all possible configurations - hence it will also create universes with the elaborate characteristics necessary for ordinary inflation to occur) together with anthropic arguments \cite{Tegmark:1997in} (which ``justify'' disregarding all universes with characteristics substantially different from ours). Such an approach may however turn out to be fallacious: eternal inflation leads to infinities, which must be regulated. However, all results turn out to depend on the regulator, \ie eternal inflation does not make regulator-independent predictions (this is the so-called measure problem of inflationary cosmology, see the contributions by Freivogel \cite{Freivogel:2011eg} and Kleban to this focus issue). Hence one may suspect that eternal inflation is not a consistent theory \footnote{Albrecht \cite{Albrecht:2009vr} and Banks \cite{Banks:2008ep} have suggested that holography may prevent eternal inflation from occurring.}, and we will adopt this point of view in this article. In this respect, it certainly seems worthwhile to examine alternative dynamical models for the early universe. In this article, we will discuss models that are based on a cosmic bounce, \ie models in which the current expanding phase of the universe was preceded by a contracting phase. We will see that such models, in particular the {\it ekpyrotic} \cite{Khoury:2001wf} and {\it cyclic} \cite{Steinhardt:2001st} models of the universe, may provide a framework that can explain the ``initial'' conditions of the universe dynamically, without recourse to anthropic arguments. Moreover, these models make rather distinctive predictions for upcoming cosmological observations, which we will discuss.

The original ekpyrotic model was inspired by string theory \cite{Khoury:2001wf}, in particular by Ho\v{r}ava-Witten theory \cite{Horava:1995qa,Horava:1996ma}. We start by reviewing a simplified version of the string theoretic setup in question (section \ref{collidingbranes}), and build up the logic that leads to a cyclic universe from there. In brief, the logic goes as follows: the big bang may be seen as a collision of branes in higher dimensions, but where the scale factor on the branes does not shrink to zero size, only the distance between the branes goes to zero. Hence the big bang may have been a much milder, and potentially tractable, event. If so, the universe would have been in a contracting phase previously. However, a contracting universe is unstable to the buildup of anisotropies and the onset of chaotic mixmaster behavior, unless there is a matter component in the universe with ultra-stiff equation of state. This matter component is ``ekpyrotic'' matter (section \ref{sectionekpyrotic}), and it can be modeled as a scalar field with a steep and negative potential (this scalar may be the scalar parameterizing the distance between the branes, its potential would then describe an attractive inter-brane force). Surprisingly, this ekpyrotic phase not only manages to flatten and isotropize the universe, but it also amplifies quantum fluctuations which may obtain a scale-invariant spectrum via the entropic mechanism reviewed below (section \ref{sectionperts}). As the potential is steep, the scalar is strongly self-coupled, and the resulting perturbations are distinguished by large non-gaussianities of the local form. However, as for inflation, for the ekpyrotic phase to be successful, the universe must again have had the right initial conditions, so it looks like we might not have made any progress at all. This is where the idea of a cyclic universe comes in (section \ref{sectionPredictivity}): it turns out that by joining the ekpyrotic phase to the future of the current dark energy phase, it becomes manifest that the current dark energy phase sets up precisely the right conditions for a subsequent ekpyrotic phase to be successful. This then leads to the idea of a cyclic universe, in which the most recent ekpyrotic phase (during which our perturbations were generated) was also preceded by a dark energy phase etc. Towards the end of this article, we will present different models for the all-important bounce phase (section \ref{sectionbounce}) and discuss the reasons why such a cyclic framework may be powerfully predictive (section \ref{sectionPredictivity}).

\section{Colliding Branes and the Big Bang}
\label{collidingbranes}

\begin{figure}[t]
\begin{center}
\includegraphics[width=0.75\textwidth]{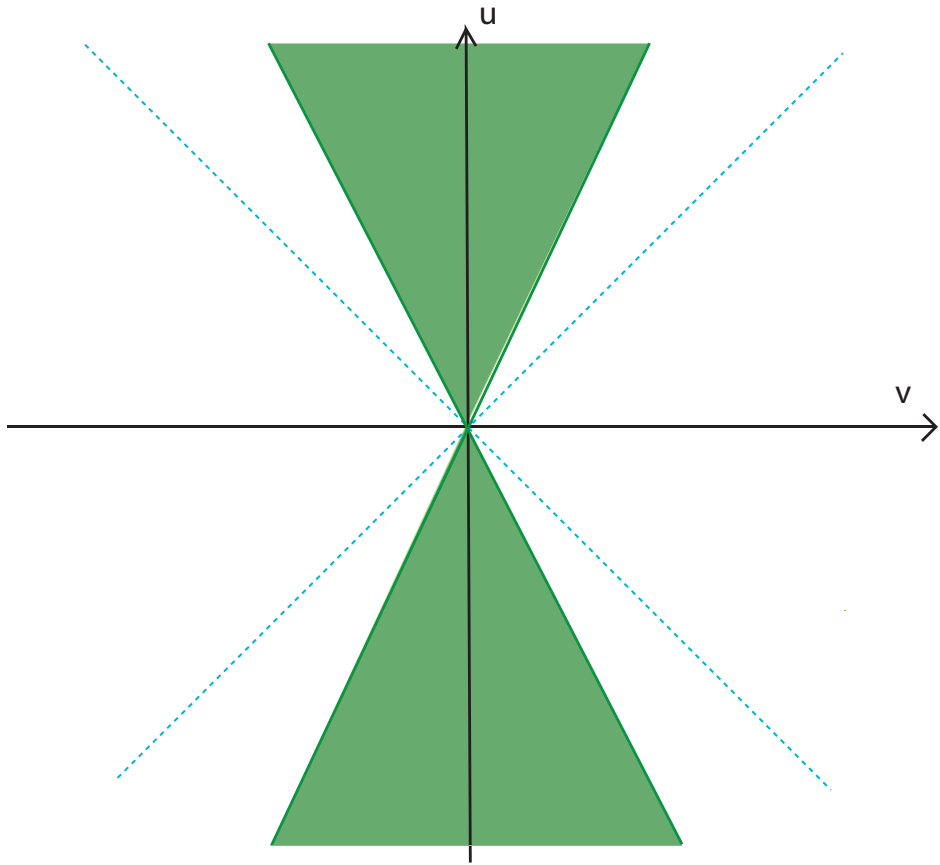}
\caption{\label{figureMilne} {\small The compactified Milne universe (the shaded region) is a useful model for a big bang caused by the collision of two boundary branes. This figure illustrates the embedding in 5-dimensional Minkowski space, the 3 ordinary spatial dimensions being suppressed and with the dashed lines delineating the light cone at the moment of the collision.
}}
\end{center}
\end{figure}

Consider pure gravity in 5 dimensions. Motivated by Ho\v{r}ava-Witten theory \cite{Horava:1995qa,Horava:1996ma}, we want to look for a solution where one spatial dimension consists of a line segment. A line segment can also be described as the quotient, or {\it orbifold}, of a circle by a $Z_2$ reflection symmetry across a diagonal of the circle. Under the action of the $Z_2$ reflection symmetry, the two endpoints of the line segment are fixed points - from the 5-dimensional point of view these define the location of two orbifold planes, which are $(3+1)$-dimensional boundaries of spacetime. We can write the metric as \be \mathrm{d} s_5^2 =
e^{-\sqrt{2/3}\phi} \mathrm{d} s_4^2 + e^{2\sqrt{2/3}\phi} \mathrm{d} y^2,
\label{5dmetric}\ee where $-y_0<y<y_0$ denotes the orbifold
coordinate and $\phi$ is the radion field parameterizing the
size of the orbifold. The $\phi$-dependent prefactor in front
of the 4-dimensional metric $\mathrm{d} s_4^2$ ensures that after
dimensional reduction we are left with the canonically
normalized Lagrangian\footnote{We use reduced Planck units $\hbar=c=1$ and
$8\pi G = M_{Pl}^{-2}=1.$} \be {\cal L} = \sqrt{-g} [\frac{1}{2}
R - \frac{1}{2}(\partial \phi)^2]. \ee Thus, the equations of motion become \be
3H^2=\frac{1}{2}\dot{\phi}^2=-\dot{H} \qquad
\ddot{\phi}+3H\dot{\phi}=0. \ee  They imply that $a
\propto e^{\phi/\sqrt{6}}$ and are solved by \be a=a_0
(-t)^{1/3} \qquad \phi=\sqrt{\frac{2}{3}} \ln(-t)+\phi_0,
\label{kinetic solution}\ee for some integration constants $a_0, \, \phi_0.$ This means that we can now plug the solution
(\ref{kinetic solution}) into the original 5-dimensional metric, to find (up to obvious
re-scalings) \bea \mathrm{d} s_5^2 &=& (-t)^{-2/3} [-\mathrm{d} t^2 + (-t)^{2/3}
\mathrm{d} x_3^2] + (-t)^{4/3} \mathrm{d} y^2 \\ &=& -\mathrm{d} T^2 + T^2 \mathrm{d} y^2 +\mathrm{d} x_3^2, \eea
where we have changed the time coordinate to $T \propto (-t)^{2/3}.$ This
model spacetime, sometimes called the compactified Milne spacetime, describes two (empty)
orbifold planes approaching each other, colliding, and receding
away from each other again.
Under the further change of coordinates $u=T \cosh y, \, v=T
\sinh y,$ the metric becomes simply that of 5-dimensional Minkowski space \be \mathrm{d} s_5^2 = -\mathrm{d} u^2 + \mathrm{d} v^2 +\mathrm{d} x_3^2, \label{metric Minkowski}
\ee which shows that this
spacetime is flat, except at the moment of collision ($T=0$), see also figure \ref{figureMilne}. From the 4-dimensional point of view, the moment of collision corresponds to $a(t)=0,$ \ie it corresponds to the big bang. However, as seen from the full 5-dimensional theory, the singularity corresponds to only the orbifold dimension shrinking to zero size, and not the other spatial dimensions. Indeed, as can be seen from (\ref{5dmetric}), the brane scale
factors are really given by $e^{-\phi/\sqrt{6}} a,$ and they are constant and non-zero at the collision. The density of matter (which is stuck on the
branes and subdominant regarding the dynamics of the collision) as well as the temperature remain finite at the collision.

This simple example shows that, in a higher-dimensional context, the big bang singularity may be much milder than in 4-dimensional general relativity, and may be tractable. We will discuss models of cosmic bounces (both singular, like this one, and non-singular) in more detail in section \ref{sectionbounce}. For now, let us just go with the idea that the big bang may not have been the beginning of the universe, and that there may have been a previous phase during which the universe contracted. This immediately leads to a potential problem, namely the wild growth of anisotropies, combined with a chaotic mixmaster big crunch, that is expected to occur in a contracting universe. In such a case, we evidently wouldn't expect a flat and isotropic universe to re-emerge from the crunch. One way to avoid this would be for the universe to be exceptionally flat at the start of the contracting phase, but this would amount once more to imposing special initial conditions - the very thing we want to explain. But there is another possibility, described in \cite{Erickson:2003zm}, which involves the addition of a new matter component. We will discuss it next.

\section{The Ekpyrotic Phase} \label{sectionekpyrotic}

For concreteness, consider a Friedmann-Robertson-Walker (FRW)
metric \be \mathrm{d} s^2=-\mathrm{d} t^2+a(t)^2\left(\frac{\mathrm{d}
r^2}{1-\kappa r^2}+r^2 \mathrm{d} \Omega_2^2 \right),
\label{FRWmetric}\ee where $a(t)$ denotes the scale factor of
the universe and $\kappa=-1,0,1$ for an open, flat or closed
universe respectively. If the universe is filled with a number
of fluids interacting only via gravity and with energy densities $\rho_i$ and constant equations of state
$w_i,$ then the equations of continuity  \be \dot{\rho_i} + 3
\frac{\dot{a}}{a}(\rho_i+p_i)=0 \label{continuity}\ee (where
dots denote derivatives with respect to time t) imply that they
will evolve according to \be \rho_i \propto a^{-3(1+w_i)}.
\label{fluidscaling}\ee The Einstein equations lead to the Friedmann
equation, which involves the Hubble parameter $H\equiv \dot a /
a$: \be H^2 = \frac{1}{3} \left(\frac{-3\kappa}{a^2}+
\frac{\rho_{m,0}}{a^3} + \frac{\rho_{r,0}}{a^4}+ \frac{\rho_{a,0}}{a^6} +
\ldots + \frac{\rho_{\phi ,0}}{a^{3(1+w_{\phi})}} \right).
\label{Friedmann} \ee The $\rho_{i,0}$s are constants giving the energy densities at scale factor $a=1$ of the various constituents of the universe: non-relativistic matter
(subscript $m$), radiation ($r$) and the energy density
associated with anisotropies in the curvature of the universe
($a$). There is also a term due to the mean curvature of space, and we assume the presence of a scalar field  $\phi.$

In a contracting universe, as $a\rightarrow 0,$ the term scaling with the largest negative power of $a$ will eventually come to dominate. And this will be the anisotropy term $\propto a^{-6},$ {\it unless} the scalar field has an equation of state $w_{\phi}>1.$ In that case, the scalar field will come to dominate and suppress the anisotropies, as well as the mean curvature, matter and radiation. Such a phase is called an {\it ekpyrotic} phase. Since the equation of state of a scalar field is given by \be w_{\phi} = \frac{p}{\rho} = \frac{\frac{1}{2} \dot\phi^{~2}-
V(\phi)}{\frac{1}{2} \dot\phi^2+
V(\phi)},
\label{scalar-eq-of-state}\ee we can see that $w_\phi > 1$ can be achieved by having a negative potential $V(\phi)$. A simple, but relevant example is the negative exponential
\be V(\phi)=-V_0 e^{-c \phi},
\label{ekpotential}\ee where $V_0$ and $c$ are constants. As mentioned previously, in string theory such a scalar could be the radion, \ie the scalar field describing the distance between the two boundary branes, and in that case a negative exponential potential can be generated by instanton effects, see {\it e.g.} \cite{Moore:2000fs,Lima:2001jc}. Once the scalar comes to dominate the dynamics, the equations of motion reduce to \be 3H^2 = \frac{1}{2}\dot\phi^2 - V_0 e^{-c\phi} \qquad \ddot\phi + 3H\dot\phi -cV_0 e^{-c\phi} =0,\ee and they are solved by \be a = (-t)^{1/\e}, \qquad \phi = \sqrt{\frac{2}{\e}} \ln
(-\sqrt{\e V_0} t), \qquad {\e}=\frac{c^2}{2}. \label{ekpyrosis-scaling} \ee The equation of state is \be w_\phi = \frac{2\e}{3}-1 = \frac{c^2}{3} -1, \ee so we need $c>\sqrt{6}$ for ekpyrosis. In realistic models (realistic in the sense of agreeing with observations), $\e$ typically turns out to be of ${\cal O}(100).$ Note that the same parameter, in inflationary models, is the ``slow-roll'' parameter and is typically of ${\cal O}(1/100).$ Here $\e$ is rather a ``fast-roll'' parameter. Note that when $\e$ is large, the potential is steep and the universe is contracting very slowly.

\section{Predictions for Observations} \label{sectionperts}

As described above, the ekpyrotic phase suppresses the mean curvature and anisotropies, and so, if this phase lasts long enough (typically a fraction of a second is already enough, see {\it e.g.} \cite{Lehners:2008vx}), the flatness problem is solved. Incidentally, the horizon problem is automatically solved too, as, during the contracting phase, there is ample time for the different regions we observe in the cosmic microwave background (CMB) to have been in causal contact. However, the ekpyrotic phase not only resolves the standard cosmological puzzles, it also generates density perturbations by amplifying quantum fluctuations \cite{Khoury:2001zk}, much like inflation, but with some crucial differences in the details. Here we simply summarize the results - the detailed derivations were recently reviewed in \cite{Lehners:2010fy}.

During the ekpyrotic phase, the universe is contracting very slowly, but the horizon $1/H \propto t \rightarrow 0$ is shrinking fast. Hence, similarly to what happens during inflation, quantum fluctuations with progressively smaller wavelengths exit the horizon and turn into classical density perturbations. Now it turns out that if there is a single scalar field, with a steep and negative potential, driving the ekpyrotic phase, then the resulting curvature fluctuations have a blue spectrum (with spectral index $n_s \approx 3$) \cite{Khoury:2001zk,Lyth:2001pf,Tsujikawa:2002qc,Creminelli:2004jg}\footnote{Right at the onset of the ekpyrotic phase, a small range of scale-invariant curvature perturbations are produced by the {\it adiabatic mechanism} \cite{Khoury:2009my,Khoury:2011ii}. However, it has so far not been possible to incorporate this mechanism into a full ekpyrotic or cyclic model of the universe.}. On large scales, these perturbations are completely subdominant (although they could make a re-appearance during the bounce phase, as discussed further on) and hence these cannot give rise to the temperature fluctuations we observe in the CMB. But if there is more than one ekpyrotic scalar present, scale-invariant curvature perturbations can be generated via the {\it entropic mechanism} \cite{Finelli:2002we,Notari:2002yc,Lehners:2007ac}. In fact, from the point of view of string theory, it is very natural to expect more than one scalar. For example, in the colliding branes setup described above, there was one scalar field describing the distance between the two orbifold planes along a fifth dimension. However, in the full string/M-theory setting, we would expect there to be 6 additional internal dimensions, for example in the shape of a Calabi-Yau manifold \cite{Lukas:1998yy}. The volume of the Calabi-Yau manifold is then described by an additional scalar, and many more scalars are required to describe all the shape deformations of the internal space. Since the inclusion of one additional scalar already entails the needed qualitative differences to the single-field case, we will focus on having two scalars for simplicity, with action \be S=\int
\sqrt{-g}[R-\frac{1}{2}(\pt\phi_1)^2
-\frac{1}{2}(\pt\phi_2)^2-V(\phi_1,\phi_2)].\ee Let us assume also that both fields get an ekpyrotic-type potential,  \be V(\phi_1,\phi_2) =-V_1 e^{-c_1 \phi_1} - V_2
e^{-c_2 \phi_2}. \label{potential2field}\ee  Then it is much
more natural to discuss the dynamics in terms of the new
variables $\s$ and $s$ pointing transverse and perpendicular to
the field velocity respectively
\cite{Koyama:2007mg,Koyama:2007ag}; they are defined, up to
unimportant additive constants which we will fix below, via \be
\s \equiv \frac{\dot\phi_1 \phi_1 + \dot\phi_2 \phi_2}{\dot\s},
\qquad s \equiv \frac{\dot\phi_1 \phi_2 - \dot\phi_2
\phi_1}{\dot\s}, \ee with $\dot\s \equiv (\dot\phi_1^2 +
\dot\phi_2^2)^{1/2}.$ For later use, we also define the angle
$\th$ of the trajectory in field space, via
\cite{Gordon:2000hv} \be \cos \th =\frac{\dot\phi_1}{\dot\s},
\qquad \sin \th = \frac{\dot\phi_2}{\dot\s}. \ee In terms of
these new variables, the potential can be re-expressed as \be
V_{ek}=-V_0 e^{\sqrt{2\e}\s}[1+\e s^2+\frac{\k_3}{3!}\e^{3/2}
s^3+\frac{\k_4}{4!}\e^2 s^4+\cdots],
\label{potentialParameterized}\ee where for exact exponentials
of the form (\ref{potential2field}), one has
$\k_3=2\sqrt{2}(c_1^2-c_2^2)/|c_1 c_2|$ and $\k_4=4(c_1^6 +
c_2^6)/(c_1^2 c_2^2(c_1^2 + c_2^2)).$ However, in the absence
of a microphysical derivation of the potential, we will simply
take $\k_3,\k_4 \sim {\cal O}(1)$ and express all results in
terms of $\k_3,\k_4.$  The
ekpyrotic scaling solution becomes \be a(t)=(-t)^{1/\e} \qquad
\s=-\sqrt{\frac{2}{\e}}\ln \left(-\sqrt{\e V_0} t\right) \qquad
s=0, \label{ScalingSolution}\ee with the angle $\th$ being
constant. Thus we can see that, up to a sign determined by convention, the field $\s$ now plays the role of the single scalar we discussed earlier, and hence this is the field that picks up a blue spectrum of curvature perturbations. The transverse field $s$, on the other hand, feels a different potential. Because of the overall minus sign in (\ref{potentialParameterized}), we can see that the potential for $s$ is actually {\it unstable} \cite{Lehners:2007ac,Tolley:2007nq}, and the motion during the ekpyrotic phase is along a ridge in the potential. This has profound consequences, which we will discuss in section \ref{sectionPredictivity}. In the present context, the instability implies that the $s$ field fluctuations $\d s$, which are entropy (or isocurvature) fluctuations, get amplified. More detailed considerations show that they obtain an amplitude $Q_s$ and a spectral index $n_s$ given by \cite{Lehners:2007ac} \bea
Q_s &\approx& (\e V_{min})^{1/2} \label{entropyamplitude}\\ \label{tilt1} n_s -1 &=& \frac{2}{\epsilon } -
\frac{\epsilon_{,N}}{\epsilon^2}, \eea where $V_{min}$ denotes the value of the ekpyrotic potential at the end of the ekpyrotic phase. The fast-roll parameter $\e$ is now allowed to vary slowly and $\mathrm{d}N\equiv \mathrm{d} \ln a.$ Hence, for $\e$ of ${\cal O}(10^2)$ (or more), the spectrum is close to scale-invariant, with the first term on the
right-hand side tending to make the spectrum blue and the second term tending to make it red\footnote{Note that during the ekpyrotic phase, $N$ is decreasing, and so must $\e$ in order for the ekpyrotic phase to come to an end eventually.}. A simple estimate, based on \cite{Khoury:2003vb}, of the natural range of
$n_s$ gives the range $0.97 < n_s < 1.02$, in good agreement with current data.

In this way the ekpyrotic phase generates nearly scale-invariant entropy perturbations. But what we are really interested in are the curvature perturbations $\zeta,$ which are the perturbations that give rise to the temperature fluctuations in the CMB. In that respect it is useful to examine the evolution equation for the (linear) curvature perturbations on large scales, given by \cite{Gordon:2000hv} \be \dot\z =
-\frac{2H}{\dot\s}\dot\th \d s. \label{zetalinear}\ee As soon as the background
trajectory bends ($\dot\th \neq 0$), the entropy perturbations
source the curvature perturbations, implying that on large scales the curvature perturbations thus created obtain the same spectrum as the entropy fluctuations (there is no wavelength dependence in (\ref{zetalinear})), with an amplitude $Q_\z$ of \cite{Lehners:2010fy} \be Q_\z^2 \approx \frac{\e V_{min}}{10^3}. \label{curvatureamplitude}\ee We will discuss the amplitude in much more detail in section \ref{sectionPredictivity}. In the most concrete models of the entropic mechanism studied in the literature (reviewed in section \ref{sectionbounce}), the indispensable bending of the trajectory occurs automatically. These models show that the bending occurs after the ekpyrotic phase has come to an end, in the phase leading up to the bounce, when the ekpyrotic potential has become unimportant and the dynamics is dominated by the kinetic energy of the $\s$ field\footnote{It was also previously envisaged that the trajectory could bend due to the instability of the potential \cite{Koyama:2007mg,Buchbinder:2007ad}. This case is now disfavored, as it leads to predictions for non-gaussianities in disagreement with observations at the $4\s$ level, see \cite{Koyama:2007if,Lehners:2010fy}.}.

The curvature perturbations created in this way are indistinguishable from those created by a phase of inflation, for example. However, at higher orders, the ekpyrotic perturbations are characterized by sizeable and specific non-gaussian signatures \cite{Buchbinder:2007at,Creminelli:2007aq,Koyama:2007if,Lehners:2010fy}. There is a simple reason for this: the potential is necessarily steep during ekpyrosis, so that the scalar fields are necessarily significantly self-coupled, leading to non-gaussian statistics for the perturbations. It turns out that these non-gaussian corrections are of the so-called ``local'' form. If the full, non-linear curvature perturbations $\z$ are re-written as a linear, gaussian piece $\z_g$, plus correction terms,
\be \z = \z_g + \frac{3}{5}f_{NL} \z_g^2 + \frac{9}{25}g_{NL}\z_g^3, \ee
then the predictions for ekpyrotic curvature perturbations generated via the entropic mechanism are \cite{Lehners:2007wc,Lehners:2008my,Lehners:2009ja,Lehners:2009qu}
\bea f_{NL} &=& \frac{3}{2}\k_3\, \sqrt{\e} + 5 \label{fNLkinetic}\\
g_{NL} &=& (\frac{5}{3}\k_4+\frac{5}{4}\k_3^2-40)\, \e.\label{gNLkinetic}\eea
With $\k_3,\k_4$ expected to be of ${\cal O}(1),$ and $\e$ typically of ${\cal O}(10^2)$ ($\e$ needs to be above about $50$ in order for the power spectrum to be in agreement with observations), we would thus expect the parameter $f_{NL}$ to be of ${\cal O}(\pm 10),$ with the sign being typically determined by the sign of $\k_3,$ and  $g_{NL}$ of ${\cal O}(-1000)$ and {\it negative} in sign. The current observational bounds are that $f_{NL} = 38 \pm 21,$ where the errors are quoted at one standard deviation \cite{Smith:2009jr}, while currently no strong constraints exist on $g_{NL}.$ It will therefore be extremely interesting to see the improved data likely to be gathered by the PLANCK satellite and ongoing and future large scale surveys, especially in light of the fact that the ekpyrotic predictions are significantly different than those of simple inflationary models.

Before proceeding to discuss the bounce in more detail, we should also discuss the predictions for primordial gravitational waves \cite{Khoury:2001wf,Boyle:2003km,Boyle:2004gv}. In fact, a quick heuristic argument is sufficient to get at the main result: at the linear level, the generation of gravitational waves depends entirely on the behavior of the scale factor. And during the ekpyrotic phase, the scale factor shrinks very slowly, so that, to a first approximation, the background is Minkowski space. This immediately implies that no significant large-scale gravitational waves are generated during ekpyrosis (otherwise gravitational waves would also be produced all around us all the time), in stark contrast to large-field inflationary models. It turns out that, at second order in perturbation theory, small-amplitude gravitational waves are sourced by the scalar density perturbations, and these are the dominant gravitational wave signal expected from an ekpyrotic phase \cite{Baumann:2007zm}, though they are unlikely to be observable in the near future. Hence a detection of large-amplitude primordial gravitational waves would rather conclusively rule out the models described here.

These predictions are interesting, because they are quite specific: reasonably large $f_{NL},$ reasonably large and negative $g_{NL}$ plus an absence of primordial gravitational waves. It is clear that inflationary models can also be constructed with exactly these predictions, hence, even if future observations remain in agreement with the predictions, no clear-cut conclusions can be drawn. However, inflationary models with exactly these predictions would have to be rather contrived and fine-tuned, and would have to be adjusted {\it a posteriori}, which would render them disfavored in the author's opinion (and according to standard results in probability theory).

\section{Cosmic Bounces: Singular or Non-Singular?} \label{sectionbounce}

In cosmological models involving a reversal from contraction to expansion, the bounce phase is the most crucial ingredient. This is the moment that gets identified with the ``big bang'', and it is the time when quantum gravity effects are likely to play a dominant role. Since we currently do not possess an experimentally supported theory of quantum gravity, all current models are necessarily rather speculative. Nevertheless, using current candidate theories, in particular string theory, we can construct and analyze various models for a bounce, and explore their consequences.

During the contracting phase, the Hubble rate is negative, while it is positive during the subsequent expanding phase of the universe. Hence, the bounce must allow the Hubble rate $H$ to increase. If we describe the dominant energy component in the universe as a perfect fluid with energy density $\rho$ and pressure $p$, then the Einstein equations imply that \be \dot{H} = -\frac{1}{2}(\rho + p).\ee Obtaining the required increase in the Hubble rate can then happen in only two ways: either $\rho + p < 0,$ \ie the null energy condition is {\it violated}, or the bounce must be classically singular with the Hubble rate instantaneously jumping from negative to positive values. Of course, especially in the second case, it becomes crucially important that the quantum corrections to the classical description be understood. We will discuss models of both types below. Before doing so, it may be useful to point out that the case of a singular bounce is {\it not} a limit of a classically non-singular bounce. In the non-singular case, the scale factor of the universe must slow down its contraction, come to a halt, then increase again. This is typically a very fine-tuned process. By contrast, in the singular case, there is typically no slowing down, as will be evident in the example described below. Hence we should not expect these two types of models to lead to equivalent physical results.

\subsection{Classically Singular Bounces}

The most interesting example of a classically singular bounce is the collision of the orbifold branes in Ho\v{r}ava-Witten theory/heterotic M-theory, a simplified version of which was already described in section \ref{collidingbranes}. Here we will elaborate on this process, and list the reasons for why this example is particularly encouraging.

Ho\v{r}ava-Witten theory describes the strong coupling limit of the 10-dimensional $E_8 \times E_8$ heterotic string theory \cite{Horava:1995qa,Horava:1996ma}. The value of the string coupling is determined by a field (the $\phi$ field of section \ref{collidingbranes}) which gets identified with the size of a new spatial dimension. As the coupling constant grows, this eleventh dimension grows too, and in this newly opened space the theory becomes that of 11-dimensional supergravity. The endpoints of the new 11th dimension consist of two orbifold fixed planes, which are endowed with an $E_8$ gauge theory each. If in addition six of the spatial dimensions are compactified on a Calabi-Yau manifold, then the theory, now usually called {\it heterotic M-theory}, contains $(3+1)$-dimensional boundary branes with realistic particle physics on them \cite{Lukas:1998yy,Lukas:1998tt,Braun:2005bw}. Restricting our attention to gravity and the Calabi-Yau volume modulus $V_{CY}$ only (these are the only fields that are relevant close to the bounce), the action is \bea S_5 &=&
\frac{1}{2\k_5^2}\int_{5d}\sqrt{-g}\left[ R-\frac{1}{2}V_{CY}^{-2}\partial_m V_{CY}\partial^m V_{CY}
-6 \a^2 V_{CY}^{-2}\right] \nn
\\ && +\frac{1}{2\k_5^2}\left\{-12\a \int_{4d,y=+1}\sqrt{-g}
\, V_{CY}^{-1} +12 \a \int_{4d,y=-1}\sqrt{-g}\,
V_{CY}^{-1} \right\} \; ,  \label{5dAction}
\eea where $\a$ denotes the boundary brane tension. The static vacuum solution to this theory is
\bea
\mathrm{d} s^2 &=& h^{2/5}(y)\,\big[A^2 \,(-\mathrm{d} \t^2 + \mathrm{d} \vec{x}^2) + B^2 \,\mathrm{d} y^2\big], \nn \\
 V_{CY} &=& B\, h^{6/5}(y), \label{domainwall} \nn \\
h(y) &=& 5\a\, y+C, \qquad -1 \leq y \leq +1
\eea
where $A$, $B$ and $C$ are integration constants.
The $y$ coordinate spans
the orbifold with fixed points at $y=\pm 1$ (we have re-scaled $y$ compared to section \ref{collidingbranes}), a negative-tension brane being located at $y=-1$ and
a positive-tension brane at $y=+1$. As we will justify below, we are interested in the situation in which the branes are moving slowly. In order to describe that case, we can simply promote the moduli $A,B,C$ to functions of (conformal) time $\t.$ Then, after the field redefinitions
\bea
a^2 &\equiv& A^2\,B\,I_{\frac{3}{5}}, \\
e^{\f_1/\sqrt{2}} &\equiv& B\,(I_{\frac{3}{5}})^{3/4}, \\
\f_2 &\equiv& -\frac{\sqrt{6}}{20}\int \mathrm{d} C\, (I_{\frac{3}{5}})^{-1}\,
\big[9\,(I_{-\frac{2}{5}})^2+16\,I_{-\frac{7}{5}}I_{\frac{3}{5}}\big]^{1/2}, \\ I_n &=& \int_{-1}^{1} \mathrm{d}y \ h^n = \frac{1}{5\a(n+1)}[(C+5\a)^{(n+1)}-(C-5\a)^{(n+1)}],
\eea
the action simplifies considerably, to \cite{Lehners:2006ir} \be S_{\mathrm{mod}} =
 \int_{4d} [-3{a'}^2 + \frac{1}{2}a^2 (\f_1^{'2} + \f_2^{'2})],
\label{ActionMSA2} \ee where $'\equiv \frac{\mathrm{d}}{\mathrm{d}\t}.$
This is simply gravity with scale factor $a$ minimally coupled to two scalars, just like the theory we used to describe the entropic mechanism in section \ref{sectionperts}, except that we have written it our here on a FRW background and in terms of conformal time. Higher-dimensional quantities are then determined in terms of the fields $a,\phi_1,\phi_2,$ for example the distance $d$ between the branes and the values of the brane scale factors $a_\pm$ at the locations of the boundaries at $y = \pm 1$: \bea
d &=& \frac{1}{3(2\a)^{1/4}}\, e^{\f_1/\sqrt{2}-\sqrt{3/2}\f_2}\,
[(1+e^{2\sqrt{2/3}\f_2})^{3/2}-|1-e^{2\sqrt{2/3}\f_2}|^{3/2}], \label{relModDistance} \\
a_{\pm} &=& (2\a)^{1/8} \,a \,e^{-\f_1/2\sqrt{2}} \,\left\{
\begin{array}{lll}
                     \left( \cosh \sqrt{2/3}\f_2 \right)^{1/4}  & \\
                     \left(-\sinh \sqrt{2/3}\f_2 \right)^{1/4}. & \\
                    \end{array} \right. \label{relModScalefactor}
\eea
The axis $\phi_2 = 0$ is special: positive values of $\phi_2$ would lead to unphysical imaginary values for the brane scale factor $a_-,$ and are thus forbidden. One can show that in the presence of matter on the boundary branes, there arises an effective repulsive potential along the $\phi_2$ axis, which causes the field space trajectory to bend in its vicinity \cite{Lehners:2006ir,Lehners:2007nb}. This is of course just what is required for the conversion of entropy into curvature perturbations described in section \ref{sectionperts}, and the example here provides a microphysical basis for the entropic mechanism.

The effective theory (\ref{ActionMSA2}) contains two scalar fields, unlike the toy model of section \ref{collidingbranes}. However, by requiring that at the brane collision the brane scale factors do not shrink to zero or blow up to infinity, we are forced to look at solutions where close to the collision $\phi_1 - \sqrt{3} \phi_2$ approaches a constant. This effectively reduces the number of scalars to one, when looking at the background evolution (for the perturbations, it remains of course important that the theory really contains two scalars). The theory (\ref{ActionMSA2}) is then solved by \be a= |2 y_0 \t|^{1/2},
\qquad e^{\f_1}= |2 y_0 \t|^{3/2\sqrt{2}}, \qquad e^{\f_2}= |4 \a
y_0 \t|^{\sqrt{3}/2\sqrt{2}}, \label{IntConst2}\ee where $2 y_0$ has the interpretation of being the velocity of the branes at the collision, as (\ref{relModDistance}) implies that $d \approx 2 y_0 \tau$ for small $\t$. Note that $y_0$ is the only parameter of the solution.

Close to the collision, the brane tension $\a$ becomes irrelevant and these solutions approach the compactified Milne solution described in section \ref{collidingbranes}, with the speed at collision being given by $2y_0.$ Their most important feature is that in the approach to the collision the spacetime that they descibe becomes increasingly flat (of course, at the moment of collision there is still a $\d$-function curvature singularity). Hence, higher-derivative corrections induced by quantum loops become more and more negligible too \cite{Lehners:2006pu}! This is similar in spirit to a conjecture due to Penrose, which states that quantum corrections are small when the Weyl curvature is small (as is the case here too) \cite{Penrose:2009zz}. So, even though this solution breaks supersymmetry\footnote{Even the exactly flat compactified Milne solution breaks supersymmetry, because the fermions are not invariant under the boost identifications that need to be performed in order for one dimension to be a line segment \cite{Khoury:2001bz}.}, the quantum corrections can be expected to be mild. This hope is given further support by the fact that the distance between the orbifold branes also determines the string coupling constant, as seen from 10 dimensions. As the branes approach each other, this coupling decreases, and it is zero at the moment of collision. Thus, during the bounce, we expect interactions to be significantly suppressed. This can be made more quantitative by studying certain aspects of the brane collision semi-classically. In particular, membranes that wind around the compact dimension are expected to become light as the size of the eleventh dimension shrinks. Thus such states must be taken into account close to the collision. Turok {\it et al.} have shown that the equations of motion for winding M2-branes remain regular at all times, and that they propagate smoothly through the collision \cite{Turok:2004gb} - see also \cite{Niz:2006ef,Niz:2007gq,Copeland:2010yr}. Instanton techniques then enable one to estimate the production of such light membranes at the collision. The result is that the density of produced membranes is finite and their gravitational backreaction is small, provided the collision speed $y_0$ remains non-relativistic. This last finding has interesting consequences regarding the amplitude of the cosmological perturbations that are produced in these models, as will be discussed in the final section below.

Eventually, one would like to be able to treat the collision of the orbifold planes of heterotic M-theory fully quantum mechanically. Unfortunately, this seems to require that a full non-perturbative definition of M-theory be found first. This drawback notwithstanding, it is interesting that all indications found so far imply that this particular example of a singular bounce may well be viable.

\subsection{Classically Non-Singular Bounces}

\begin{figure}[t]
\begin{center}
\includegraphics[width=0.75\textwidth]{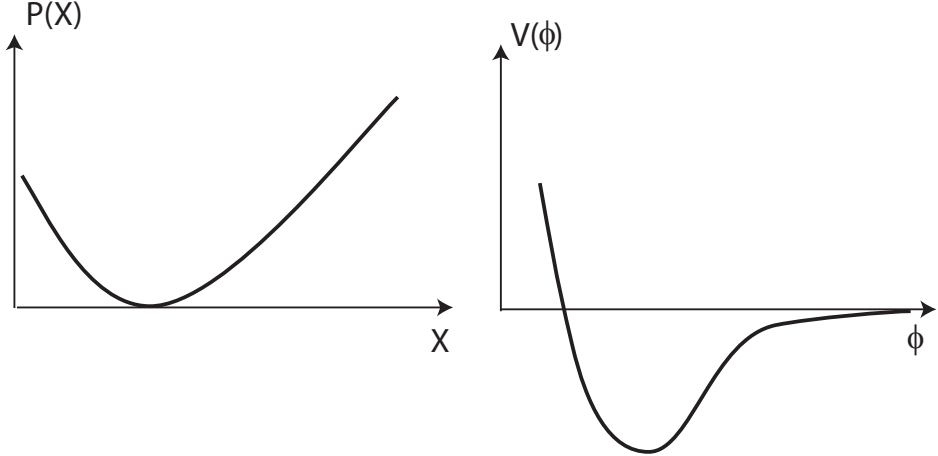}
\caption{\label{figureGhost} {\small The kinetic and potential functions for the ghost condensate bounce, as used in new ekpyrotic models.
}}
\end{center}
\end{figure}

As we saw at the beginning of this section, in order for the universe to bounce in a non-singular way, the dominant energy component in the universe must violate the null energy condition (NEC). From a model-building point of view, this is difficult to achieve, as violations of the NEC are commonly associated with various instabilities in the theory (most notably the appearance of ghosts, \ie fields with negative kinetic energy). Moreover, since the energy density scales as $a^{-3(1+w)}$ for matter with equation of state $w,$ and since violations of the NEC correspond to $w<-1,$ we can see immediately that in a contracting universe such matter becomes quickly less and less relevant. In fact, as discussed in section \ref{sectionekpyrotic}, at the end of the ekpyrotic phase all other types of matter have become vastly subdominant to the ekpyrotic scalar field. Hence, the only option seems to be that the ekpyrotic scalar itself must turn into the required NEC-violating matter \cite{Buchbinder:2007ad}. At that point, our options are currently rather limited: there exist only two types of scalar field theories in which the NEC can be violated in a stable and controlled way: ghost condensates \cite{ArkaniHamed:2003uy} and galileons \cite{Nicolis:2008in,Nicolis:2009qm,Creminelli:2010ba}.\footnote{In fact, it has recently been discovered in \cite{Khoury:2011da} that both types of theories are closely related: they are virtually identical when one considers time dependence alone, the main differences residing in spatial gradient terms.} We will mainly focus on the most studied case, that of a ghost condensate bounce, which has been used in ``new ekpyrotic'' models \cite{Buchbinder:2007ad}, see also \cite{Creminelli:2007aq,Lin:2010pf}. It works as follows: suppose that, as the kinetic energy for the scalar field $\phi$ becomes very small, its kinetic term becomes non-canonical. In fact, if we denote $-\frac{1}{2m^4}(\pt\phi)^2 \equiv X,$ so that the ordinary kinetic term in the Lagrangian is simply $X,$ then we imagine that, for small $X$, the Lagrangian (which we denote by ${\cal L}=P(X)$) is of the form \be {\cal L}_{small \, X} = P(X)_{small \, X} = M^4(-X+X^2),\ee \ie it has a local minimum. Here $m$ and $M$ are two mass scales, which we will discuss in more detail below. The full kinetic function is sketched in figure \ref{figureGhost}.
When the scalar reaches the minimum, the vacuum shifts, in the sense that the scalar develops a time-dependent vacuum expectation value that is growing linearly with time, $\langle \phi \rangle  = m^2 t.$ The Lagrangian, perturbed to quadratic order around this time-dependent vev, is proportional to
\be {\cal L} \propto (2XP_{,XX} + P_{,X})(\dot{\d\phi})^2 - P_{,X} (\nabla \d\phi)^2. \label{perturbedghost} \ee
Amazingly, the perturbations in the vicinity of the minimum ($P_{,X} \approx 0$) are not ghost-like, despite the $-X$ term in the Lagrangian. The energy density and pressure are given by \be \rho = 2P_{,X} X - P, \qquad p = P,\label{energypressure}\ee so that $\rho + p = 2P_{,X}X.$ Thus, at the minimum, the ghost condensate is right on the verge of violating the NEC, its equation of state being that of a cosmological constant $p=-\rho.$ But if the field is pushed to even smaller $X$ values, where $P_{,X}<0,$ then it violates the NEC, {\it without the appearance of ghosts}, and can cause the universe to revert smoothly from contraction to expansion.

For such a scenario to work, a number of conditions must be met: first, there needs to be a steeply rising potential to slow the scalar down after the end of the ekpyrotic phase, in order for the ghost condensate phase to start - see figure \ref{figureGhost} for a sketch of the shape of the potential that is required. In fact, as described in \cite{Buchbinder:2007ad}, the potential must rise roughly as $V \propto -\frac{1}{m^2}\phi$ for consistency.

Secondly, even though there are no ghosts, there are still gradient instabilities when the NEC is violated, as is apparent from the last term in (\ref{perturbedghost}). However, as shown in \cite{Nicolis:2009qm,Buchbinder:2007ad}, these are harmless as long as the bounce is fast (on the order of one e-fold), as there is then no time for the instability to grow. But having a fast bounce requires the parameters of the model, in particular the mass scales $m$ and $M,$ to be rather finely tuned. It turns out that, for an ekpyrotic phase with fast-roll parameter $\e,$ lasting for $N_{ek}$ e-folds and ending at a potential value of $V_{min},$ the mass scales must satisfy the hierarchy \be m^4 e^{2N_{ek}} \ll |V_{min}| \ll \e M^4. \label{parameterrange}\ee In words, the effective field theory cut-off scale $M$ must be at or above the scale of the ekpyrotic potential, with the scale $m$ of the ghost condensate vev many orders of magnitude below. Such a tuning does not seem unreasonable.

There is one further requirement that the ghost condensate model must satisfy, and it stems from a potential problem that has recently been discovered by Xue and Steinhardt, and which actually threatens to invalidate all non-singular models of a bounce \cite{Xue:2010ux}. The problem arises from an unlikely source, namely an apparently doubly subdominant cosmological perturbation mode. Let us temporarily neglect the entropy perturbations that are thought to source the primordial density fluctuations in the present model. Then, as discussed at the beginning of section \ref{sectionperts}, the dominant perturbations have a blue spectrum and ultimately provide a negligible contribution to the observed temperature fluctuations in the CMB. In slightly more detail, the comoving curvature perturbation $\z$ obeys the equation of motion (in Fourier space and conformal time) \be \z_k'' + 2\frac{z'}{z}\z_k' + k^2 \z_k = 0,\ee where $z\equiv a\sqrt{2\e/c_s^2}.$ $c_s$ denotes the speed of sound of the fluctuations, and can be read off from (\ref{perturbedghost}), \be c_s^2 = \frac{P_{,X}}{2XP_{,XX} + P_{,X}}.\label{speedofsound}\ee On large scales, \ie for small $k,$ the equation of motion is solved by \be \z_k = \frac{c_1}{\sqrt{k}} + c_2 \sqrt{k}\int \frac{\mathrm{d}\t}{z^2},\label{zetasolution}\ee where $c_1,c_2$ are constants of ${\cal O}(1),$ as can be seen by matching to the Minkowski vacuum in the far past. The first mode is the one with the blue spectrum $n_s = 3,$ and which can be neglected. The second mode has an even bluer spectrum, $n_s = 5,$ and is typically not even mentioned. But it is instructive to re-write the integral in (\ref{zetasolution}) as \be \frac{\z_k^{int}}{c_2 \sqrt{k}} \equiv \int \frac{\mathrm{d} \t}{z^2} = \int \frac{\mathrm{d} t \, c_s^2}{2 a^3 \e} = \int \frac{\mathrm{d}t \, c_s^2}{3a^3 (1+w)}.\ee During the bounce, we must have $w<-1,$ and so it looks like the integral might blow up in the approach to the bounce phase, which would signal a breakdown of perturbation theory. However, the speed of sound also goes to zero as $w$ reaches $-1,$ and so the integral is finite. It is nevertheless dominated by the contribution close to the bounce phase, and can be evaluated using (\ref{energypressure}), (\ref{speedofsound}) and the approximation of neglecting the friction term in the equation for $X,$ $\dot{X} \approx - V_{,\phi}\sqrt{X},$ \be \frac{\z_k^{int}}{c_2 \sqrt{k}} = \int \frac{2XP_{,X}-P}{2a^3X(2XP_{,XX}+P_{,X})}\frac{\mathrm{d}X}{(-V_{,\phi})\sqrt{X}} \approx \frac{-1}{V_{,\phi}\sqrt{X_{bounce}}}. \ee Since the speed of the scalar at the bounce $\sqrt{X_{bounce}}$ is necessarily very small, we can see that in the approach to the bounce, this initially doubly subdominant term gets massively amplified, and threatens to dominate over all other perturbation modes. As shown in \cite{Xue:2010ux}, the ratio of this mode to the curvature perturbation $\z_s$ produced by the entropic mechanism is \be \frac{\z^{int}}{\z_s} \approx \frac{-1}{V_{,\phi}}e^{N_{ek} - 2 N_k},\ee where $N_k$ denotes the number of e-folds of ekpyrosis remaining after mode $k$ of $\z^{int}$ has exited the horizon. For modes of observational interest $N_k \approx (20-30),$ while $N_{ek}>60,$ so for many models of non-singular bounces this effect will be problematic. However, as we saw above, in the present model the bounce must be fast in order to avoid gradient instabilities, and this means that the potential must be steep, $|V_{,\phi}| \approx 1/m^2 \gg 1.$ In fact, the consistency relation (\ref{parameterrange}) implies that for the model discussed here, $m^2 e^{N_{ek}}$ is always small, and so the doubly blue mode remains subdominant. Hence, the condition that the bounce be {\it fast} avoids the onset of all known potential instabilities.

From a model-building point of view, it is very satisfying to see that a consistent non-singular bounce model can be constructed\footnote{See the review \cite{Novello:2008ra} for other attempts at constructing bounce models.}. What is not so clear yet is whether such a model could also arise from string theory. There exist some general arguments to the contrary \cite{Adams:2006sv,Kallosh:2007ad}, but history has shown that no-go arguments have a tendency to be overcome. In a recent effort to clarify this question, the ghost condensate model has been supersymmetrized \cite{Khoury:2010gb}, which may help in elucidating the connection to string theory. It would also be interesting to see if non-singular bounce models can be constructed using the closely related galileon theories. These theories do not suffer from gradient instabilities, and so might lead to more general bounce models. The relation of these models to string theory is equally unsettled at present. They are thought not to arise in string theory by the arguments of \cite{Adams:2006sv}, but on the other hand, they can be derived, very suggestively, from the dynamics of branes \cite{deRham:2010eu}. They have also been supersymmetrized recently \cite{Khoury:2011da}, and so these questions may find a clarification soon.

\subsection{Nonperturbative Bounces}

The AdS/CFT correspondence is thought to provide a non-perturbative definition of a theory of quantum gravity. This makes one wonder whether, in this framework, one can finally tackle the question of cosmological singularities in a full quantum mechanical way, and study bounces non-perturbatively. In this vein, Craps {\it et al.} have attempted to construct models of a crunch of Anti-de Sitter space \cite{Turok:2007ry,Craps:2007ch,Craps:2009qc}, which are described in detail in his contribution to the present volume. Hence, we will limit ourselves to just a few remarks here, to point out the main differences with the colliding branes bounce discussed above: the AdS crunch corresponds to a high curvature crunch, and thus can be expected to lead to overproduction of particles at the crunch, with subsequent rapid re-collapse. Numerical studies in the context of a 5-dimensional model confirm this expectation \cite{Battarra:2010id}, but there is the hope that a better behavior of the coupling constant in the dual conformal field theory for the 4-dimensional case might evade this problem. Also, in the models studied so far, there is no ekpyrotic matter present, and thus there is no matter component that can halt the rapid growth of anisotropies. Thus, the models in question may describe the quantum version of chaotic mixmaster collapse. This in itself will be very interesting, but at the same time it will also be of interest to see if the framework can be extended to low-curvature crunches more like that of compactified Milne space.

\subsection{Reheating}

In all models of a cosmic bounce, the ordinary radiation and matter content of the universe is thought to arise at the bounce itself. For singular bounces, one expects the brane collision to be slightly inelastic, so that a small part of the collision energy will be converted into radiation and matter \cite{Steinhardt:2001st}. Note that in that case the scalar fields do not disappear by decaying into radiation and matter, as is typically assumed in inflation. Rather, the scalar fields become cosmologically relevant again at later times, when they act as quintessence, and their non-decay is essential in the construction of a cyclic model of the universe, as detailed in the next section. For non-singular bounces on the other hand, the reheating mechanism most discussed so far involves the decay of the field governing the bounce, {\it e.g.} the ghost condensate field, into radiation and matter just after the bounce \cite{Buchbinder:2007ad}. Such a mechanism is most suitable for single-bounce models of the universe, but appears more difficult to incorporate into a cyclic model.

It seems fair to say that the topic of reheating has not been studied in sufficient detail and generality in the context of cosmic bounces; this is certainly an area that deserves further research.

\section{Cosmic Cycles, the Phoenix and Predictivity} \label{sectionPredictivity}

By having a contracting phase before the current expanding phase, have we not simply pushed the question of initial conditions back to an earlier time, in much the same way as the theory of inflation does, though with the added complication of a bounce? Then all we have gained is a new mechanism for generating cosmological perturbations (though we already had one with inflation), but no answers to any other basic cosmological questions. However, imagine that we link up alternating contracting and expanding phases into a cyclic universe, could it then be that one cycle prepares the initial conditions for the next one? Or, assuming that there was some original big bang-like event, but which did not produce a universe like ours - could subsequent cycles evolve more and more into universes like ours? If so, such an attractor framework might be highly predictive, much more so than a theory which produces all types of universes, but with no clear preference for one like ours\footnote{This kind of argument can be made precise, see for example \cite{Neal:2006py}.}. The cyclic model which we will discuss now is an attempt to build precisely such a model.

\begin{figure}[t]
\begin{center}
\includegraphics[width=0.75\textwidth]{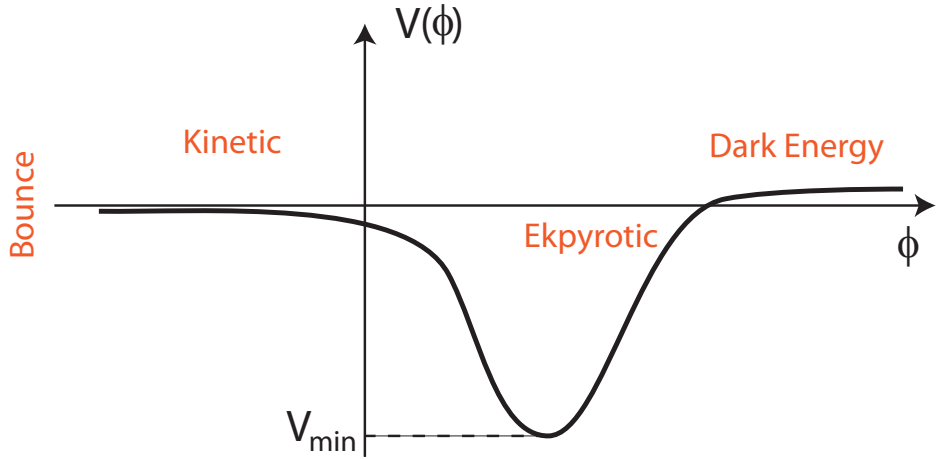}
\caption{\label{figurecyclic} {\small The effective potential in the cyclic universe, with the different phases of evolution.
}}
\end{center}
\end{figure}

The cyclic universe, first proposed by Steinhardt and Turok \cite{Steinhardt:2001st}, posits that the current dark energy phase that we find ourselves in will be followed in the distant future by a contracting ekpyrotic phase, which will lead to a bounce and a new phase of radiation and matter domination, followed once more by dark energy, and so forth. As an effective theory, this can be described by a scalar field moving in the potential shown in figure \ref{figurecyclic}. The figure shows the potential only in one direction, and one should keep in mind that other, transverse, directions can also be relevant: in particular, during the ekpyrotic phase, there is a transverse unstable potential, as described in section \ref{sectionperts}, which will play an important role here.

The cyclic model is formulated as an effective theory, without specifying a precise microphysical origin, although the most precise formulation to date is based on the colliding branes picture stemming from heterotic M-theory\footnote{It is at present an open question whether a consistent cyclic model of the universe can also be constructed using a non-singular bounce of the ghost condensate type, for example.}. It is useful to briefly go over the different phases of the model, and to follow the evolution of the scale factor and the Hubble parameter \cite{Erickson:2006wc}. Let us start with the current dark energy phase, when the scalar field takes its largest values. During this phase, the Hubble parameter is roughly constant and the universe grows by $N_{de}$ e-folds as the scalar field is very slowly rolling to smaller field values. Eventually the potential becomes negative, the field speeds up, the universe starts to contract again and a new ekpyrotic phase takes place. From section \ref{sectionekpyrotic} we know that the scale factor now remains roughly constant, while the Hubble parameter grows by $e^{N_{ek}}=V_{min}^{1/2}/V_0^{1/2},$ where $V_0$ denotes the current dark energy density. The ekpyrotic phase comes to an end when the scalar reaches the bottom of the potential at $V_{min},$ and is followed by the bounce phase. If we assume the colliding branes model of the bounce, the dynamics will be dominated by the kinetic energy of the scalar, with the scalar shooting off to $-\infty$ and bouncing back after the brane collision. By the end of the bounce/kinetic phase, the universe has grown by the modest amount $V_{min}^{1/6}/T_r^{2/3},$ where $T_r$ denotes the temperature to which the universe reheats at the brane collision. Meanwhile, the Hubble parameter has shrunk by a factor $T_r^2/V_{min}^{1/2}.$ After the bounce, the scalar field flies back over the potential well, slows down due to Hubble friction and comes to a halt on the plateau at large field values. Right before the dark energy phase, while the energy density of the universe is still dominated by radiation and matter, the universe grows by a factor of about $T_r/T_0,$ with the Hubble parameter shrinking by $T_0^2/T_r^2,$ where $T_0=V_0^{1/4}$ is the current temperature of the universe. Thus, putting everything together, the Hubble parameter returns to its starting value after one cycle (as do all such locally measurable quantities), while the scale factor grows by the large amount of \be \Bigg(\frac{V_{min}^{1/4}}{T_r}\Bigg)^{2/3}
\Bigg(\frac{T_r}{T_0}\Bigg) e^{N_{de}},\label{netgrowth}\ee over the course of one cycle. This net growth is essential to the viability of the cyclic universe. For one, it solves a puzzle about the second law of thermodynamics, which asks how a cyclic universe is possible, given that the entropy must increase all the time. The resolution here is that, although the total entropy does increase continually, the entropy {\it density} gets diluted to the same small value again after each cycle.

We should now go back to the ekpyrotic phase, and address the instability of the potential that is necessary for the generation of perturbations. Cyclic universe and instability are two concepts that do not seem to go very well together {\it a priori}. However, this instability may actually lie at the heart of the high level of predictivity of the model. As described in section \ref{sectionperts}, during the ekpyrotic phase the background trajectory of interest is the one that rolls unstably along a ridge in the potential, with the potential falling off steeply both left and right. For slightly different initial conditions, the ekpyrotic phase will be of short duration: the trajectory will veer off to the side, ekpyrosis and the associated flattening of the universe will end prematurely, and the corresponding region of the universe will collapse in a chaotic mixmaster crunch. At the end of such a crunch, we will conjecture that a gravitationally collapsed region remains, in the form of a black hole presumably, so that this region will stop growing. Only a region with very special field values at the start of the ekpyrotic phase will make it all the way down the ridge, through the bounce and on to the next expanding phase. However, that region, unlike all the other collapsed regions, will grow by the enormous factor (\ref{netgrowth}) until the start of the next ekpyrotic phase. Moreover, for $N_{de}>60,$ it turns out that this newly expanded universe is bigger than its originator universe one cycle before \cite{Lehners:2008qe}. Thus, as long as the dark energy phase lasts sufficiently long (sufficiently long meaning at least 600 billion years), the cyclic universe grows from cycle to cycle and {\it selects} regions of the universe with just the right initial conditions to produce a large, flat and habitable space during the next cycle. In this context, it is wrong to say that the ekpyrotic phase requires special initial conditions. Rather, this {\it phoenix universe} \cite{Lehners:2008qe,Lehners:2009eg}, reminiscent in spirit of the phoenix universe discussed by Lemaitre \cite{Lemaitre} and Dicke, produces a vast range of ``initial'' conditions before the onset of the ekpyrotic phase\footnote{Note that, given the assumed shape of the potential, this range of ``initial'' conditions does not however include conditions which would allow eternal inflation to occur. If this were the case, we would of course face the measurement problem once again.}. Almost the entire universe will then collapse irredeemably, except for some small regions, which are necessarily exceptionally flat, homogeneous, isotropic but with small quantum generated density perturbations, growing into the kind of universe we see around us. And should the universe that is produced in such a way happen to be small (compared to the presently observed Hubble volume, say) then this is hardly an obstacle, as each subsequent cycle will make it bigger and bigger. Unlike eternal inflation, the phoenix universe renders the vast majority of space habitable. Note also that it is not necessary for the cycles to have occurred indefinitely into the past. A beginning of the universe, via a tunneling event for example, is entirely conceivable in this framework, as long as such an event produces one region with the right conditions for the ekpyrotic phase to proceed successfully. Whether the universe had a beginning or not is unfortunately difficult to establish experimentally in this framework. For better or worse, the instability implies a near total isolation from even the immediately preceding cycle, with only the quantum generated fluctuations surviving the bounce. Any small excess energy present at the onset of the last ekpyrotic phase in the region that is supposed to evolve into our observable universe would automatically cause the corresponding field trajectory to fall off the potential ridge and prevent that region from making it through the bounce. On the positive side, this means that there is no paradox in us not having found messages across the sky from far-advanced civilizations of the last cycle(s).

\begin{figure}[t]
\begin{center}
\includegraphics[width=0.75\textwidth]{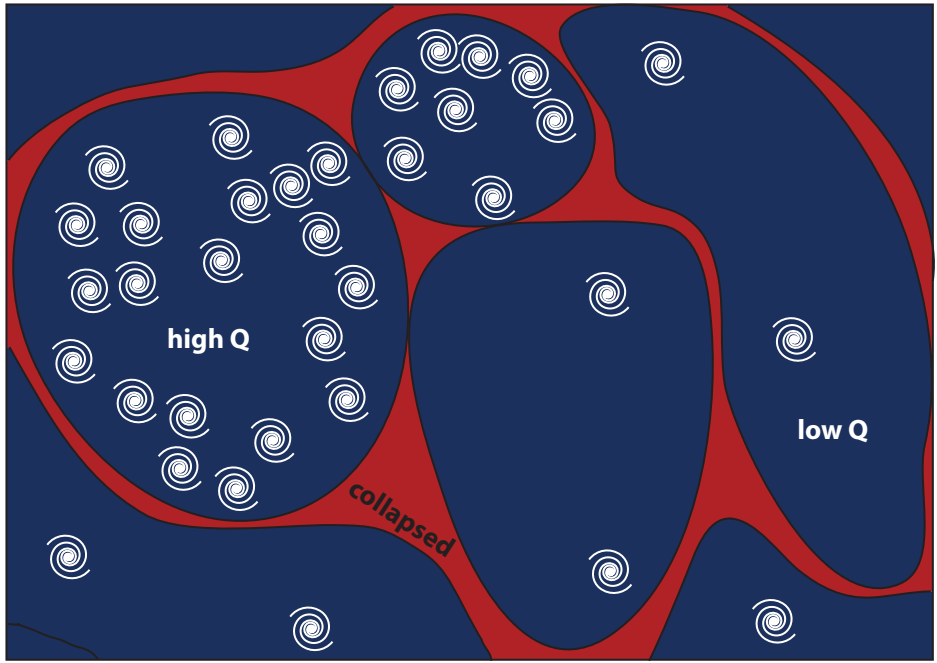}
\caption{\label{figurePhoenix} {\small The global structure of the cyclic universe.
}}
\end{center}
\end{figure}

As speculative as some of these discussion may sound, the cyclic universe described above leads to concrete pre- and postdictions, which can be confronted by experiment. The most important predictions, regarding non-gaussianities in the CMB and gravitational waves, were described in section \ref{sectionperts}. But there is also an important postdiction, concerning the amplitude $Q$ of the primordial density perturbations \cite{Lehners:2010ug}. In other models of the early universe, notably inflationary models, this quantity is always adjusted by hand to fit the observed value of $Q_{obs}\sim 10^{-5},$ even though the observed value is typically not favored from a theoretical point of view. By contrast, the cyclic model (with a bounce modeled by colliding branes) {\it dynamically selects} regions of the universe with the observed value of $Q$! This is entirely analogous in spirit to the selection of flat regions of the universe discussed in the previous paragraph. Since this is a significant result, it is worth deriving it in a little bit more detail. The ekpyrotic phase evolves according to
the scaling solution (\ref{ScalingSolution}), or, using conformal time, \be
a=(-\tau)^{1/\e}, \quad \frac{\mathrm{d}\phi}{\mathrm{d}\tau} = \sqrt{2}/({\sqrt{\e}\,\t}),
\quad V = -\e {\cal H}^2, \ee with the conformal Hubble rate ${\cal H} \equiv (\frac{\mathrm{d}a}{\mathrm{d}\tau})/a.$ The subsequent kinetic phase is characterized by \be a = (-
\t)^{1/2}, \qquad \frac{\mathrm{d}\phi}{\mathrm{d}\tau}= {\sqrt{3}}/({\sqrt{2}\,\t}). \ee
Matching
the conformal Hubble rates in the two phases at the time $\t_k$ that the kinetic phase begins, implies \be -2\t_k =
\sqrt{{\e}/{V_{min}}}.\label{matchingHubble}\ee Recalling also from the discussion below Eq. (\ref{IntConst2}) above that the inter-brane distance is given by \be d = 2y_0 |\t_k|,\ee we get the relation $V_{min} = \e y_0^2/d^2.$ An important additional information is that the extension of the solution (\ref{IntConst2}) to the ekpyrotic phase implies that the inter-brane distance is approximately constant during that phase, with the value $d$ given above (see \cite{Lehners:2006ir,Lehners:2010ug}). Hence, the amplitude $Q_{\z}$ of the curvature perturbations generated by the entropic mechanism, given above in (\ref{curvatureamplitude}), can be re-expressed very simply as \be Q_\z \approx \frac{\e y_0}{10^{3/2}d} \approx \frac{y_0}{d},\ee where we have taken $\e \sim {\cal O}(10^2)$ both because this is consistent with observations of the spectral index, and because, from a theoretical point of view, the lowest allowed values for $\e$ correspond to the least amount of fine-tuning in the potential. For $d$, we can assume the value $d\approx 10^{3.5}$ suggested by Ho\v{r}ava-Witten theory, and which implies the observed value of Newton's constant. The crux of the argument comes from considering $y_0$, the brane collision velocity. As discussed in section \ref{sectionbounce}, it cannot be arbitrarily high, as this would lead to an overproduction of matter (wrapped M2-branes) at the brane collision, with a subsequent rapid collapse of the corresponding region. In other words, if $y_0$ is relativistic in some region, then that region undergoes gravitational collapse at the bounce, and becomes irrelevant regarding the large-scale evolution of the cyclic universe in much the same way as those regions with the wrong initial conditions for ekpyrosis described above. Thus, the vast majority of the universe must have $y_0 \lesssim 0.1,$ or \be Q_\z \lesssim 10^{-4.5}. \label{Qupperbound}\ee Regions in which the brane collision velocity is higher lead to a larger $Q_\z$ and thus to more structure, making it highly likely to find oneself in a region of the universe with an observed value of $Q$ as close as possible to the upper bound (\ref{Qupperbound}), see also figure \ref{figurePhoenix}. It is certainly intriguing that the observed value of $Q$ agrees so closely with this postdicted value!

To conclude, it seems worthwhile to dwell for a little longer on the main theme discussed in this last section, namely the contrast between the instability of the ekpyrotic phase and the net expansion over the course of each cycle. The discarding of large regions of the universe at the end of each cycle provides a type of dissipation, which is just what is needed to have attractor behavior of the model. Hence, perhaps somewhat counter-intuitively, it is precisely the interplay between {\it fragility} and {\it amplification} that lends the model its high predictive power. If it was easy to make all sorts of large, reasonably flat universes, with all sorts of density perturbations, all allowing for the formation of gravitationally bound structures, it would be very difficult to understand the broad features of our universe. We would have to conclude that our part of the universe is as it is because of sheer coincidence. Before accepting such a conclusion though, we should test those models in which the specialness of our universe is not due to mere historical accident, but due to mathematical and physical necessity.

\section*{Acknowledgments}

I would like to thank Thorsten Battefeld, Justin Khoury, Jens Niemeyer, Burt Ovrut, Paul Steinhardt, Neil Turok, Bingkan Xue for illuminating discussions on the topics discussed in this article. The author gratefully acknowledges the European Research Council's support in the form of a Starting Grant.

\bibliography{CosmicBounces}

\end{document}